\documentclass[aps,prl,twocolumn,groupedaddress,preprintnumbers]{revtex4}

\usepackage{color}
\usepackage{graphicx}
\usepackage{amssymb, amsmath, amsfonts}

\begin{document}

\title{Phase-change chalcogenide glass metamaterial}

\author{Z. L. S\'{a}mson}
\email{zls@orc.soton.ac.uk}
\author{K. F. MacDonald}
\affiliation{Optoelectronics Research Centre, University of Southampton, Southampton, SO17 1BJ, UK}
\author{F. De Angelis}
\affiliation{NanoBioScience Lab, Istituto Italiano di Tecnologia, Via Morego 30, I16163 Genova, Italy}
\affiliation{BIONEM Lab, University of Magna Graecia, viale Europa, I88100 Catanzaro, Italy}
\author{K. Knight}
\author{C. C. Huang}
\author{D. W. Hewak}
\affiliation{Optoelectronics Research Centre, University of Southampton, Southampton, SO17 1BJ, UK}
\author{E. Di Fabrizio}
\affiliation{NanoBioScience Lab, Istituto Italiano di Tecnologia, Via Morego 30, I16163 Genova, Italy}
\affiliation{BIONEM Lab, University of Magna Graecia, viale Europa, I88100 Catanzaro, Italy}
\author{N. I. Zheludev}
\homepage{http://www.metamaterials.org.uk}
\affiliation{Optoelectronics Research Centre, University of Southampton, Southampton, SO17 1BJ, UK}

\date{\today}

\begin{abstract}
Combining metamaterials with functional media brings a new dimension to their performance. Here we demonstrate substantial resonance frequency tuning in a photonic metamaterial hybridized with an electrically/optically switchable chalcogenide glass. The transition between amorphous and crystalline forms brings about a 10\% shift in the near-infrared resonance wavelength of an asymmetric split-ring array, providing transmission modulation functionality with a contrast ratio of 4:1 in a device of sub-wavelength thickness.
\end{abstract}

\maketitle

With the potential to impact substantially across fields ranging from telecommunications and defence to renewable energy and healthcare, metamaterials – nanostructured media with extraordinary electromagnetic properties not found in nature – have become the subject of intense investigation~\cite{Ozbay2008}. A remarkable array of new fundamental physical phenomena and functionalities have been demonstrated, for example: asymmetric transmission~\cite{Fedotov2007}; optical magnetism~\cite{Enkrich2005}; optical activity without chirality~\cite{Plum2009}; negative refractive index~\cite{Smith2004}; induced transparency~\cite{Papasimakis2009}; super-lensing~\cite{Pendry2000, Zhang2008}; and cloaking~\cite{Pendry2006, Li2008}.

The functionality of metamaterials, typically comprising an ensemble of identical metallic nanostructures embedded in or supported by a dielectric, is underpinned by a resonant response characteristic related to the structural dimensions and material composition. Indeed, it has been found that exceptionally narrow, high-quality resonances (`trapped-modes': weakly coupled current oscillations exhibiting low radiation losses) can be excited in asymmetric split-ring (ASR) resonators~\cite{Fedotov2007a}. The ability to adaptively tune or reversibly switch the frequency of such resonances would radically enhance metamaterial functionality in numerous potential applications and simultaneously open up entirely new possibilities.

Frequency shifts have been achieved in microwave and terahertz metamaterial composites incorporating semiconductors, transition metal oxides and varactors as active media/elements~\cite{Driscoll2009, Chen2008, Gil2004}, and it has recently been shown that the insulator-metal transition in vanadium dioxide allows thermally-controlled infrared resonance tuning in Ag/VO$_2$ hybrid structures~\cite{Dicken2009}. Here, we demonstrate non-volatile resonant frequency switching in an active near-infrared metamaterial comprising a gold ASR array hybridized with a bistable chalcogenide glass.

The chalcogenides - compounds containing heavier Group 16 elements (S, Se, Te) - are a remarkable family of materials, displaying an extensive range of technologically-relevant optical, electronic, thermal and mechanical effects~\cite{Popescu2006}. They have been in widespread commercial use for many years in re-writable optical discs (where data is routinely encoded at more than 200~Mbit/s)~\cite{Wuttig2007} and appear set to form the basis of next-generation electronic memories~\cite{Welnic2009}. They are also being intensively studied for solar cell~\cite{Repins2008}, infrared~\cite{Shaw2001} and nonlinear optical applications~\cite{Sanghera2008}. Their threshold switching properties - the fact that they can be rapidly and reversibly converted between amorphous and crystalline phases with markedly different electromagnetic properties - have made them the material of choice for optical and electronic memory applications. We now show that this functionality can be brought to bear in the metamaterials domain to create active devices.

\begin{figure}[!b]
\includegraphics*[width=\columnwidth]{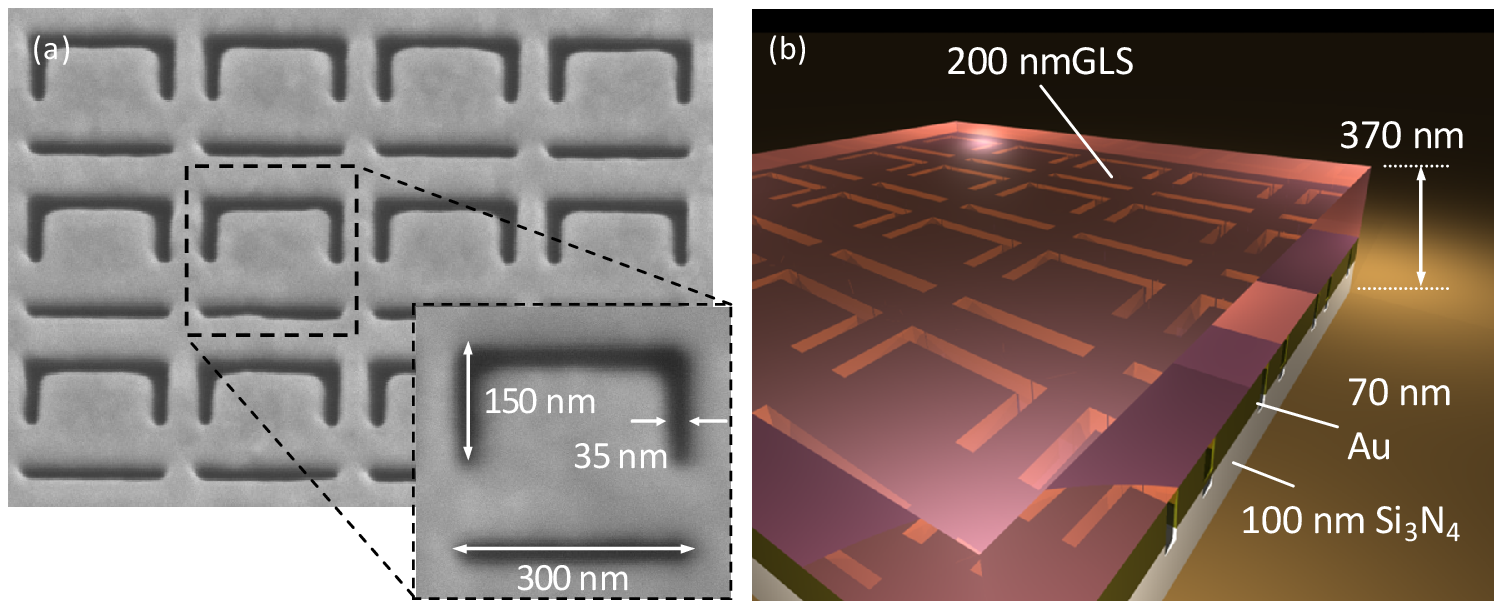}
\caption{(a) Scanning electron microscope image of part of the gold asymmetric split-ring (ASR) resonator array before deposition of the gallium lanthanum sulphide (GLS) chalcogenide film. The inset shows the structural dimensions of an individual unit cell. (b) Artistic rendition of the gold-ASR/GLS hybrid metamaterial structure, showing a cross section of the device, which has an overall thickness of 370~nm. [Image credit: G. Adamo]}
\label{sample}
\end{figure}

The present study employed a gold negative ASR array metamaterial hybridized, as illustrated in Fig.~\ref{sample}, with gallium lanthanum sulphide (GLS) - a thermally stable semiconducting chalcogenide glass with a bandgap of 2.6~eV and a band edge across the visible spectral range.

Each 375$\times$375~nm unit cell of the 30$\times$30~$\mu$m metamaterial array contained a 300$\times$300~nm square ASR with a 35~nm line width, fabricated by focused ion-beam milling in a 70~nm gold layer supported on a 100~nm silicon nitride (Si$_3$N$_4$) membrane. An amorphous GLS sputtering target was prepared by melt quenching and annealing a mixture of high-purity gallium sulphide and lanthanum sulphide precursors and a 200~nm GLS film (monitored by quartz crystal microbalance) was sputtered onto the ASR array.

\begin{figure}
\includegraphics*[width=\columnwidth]{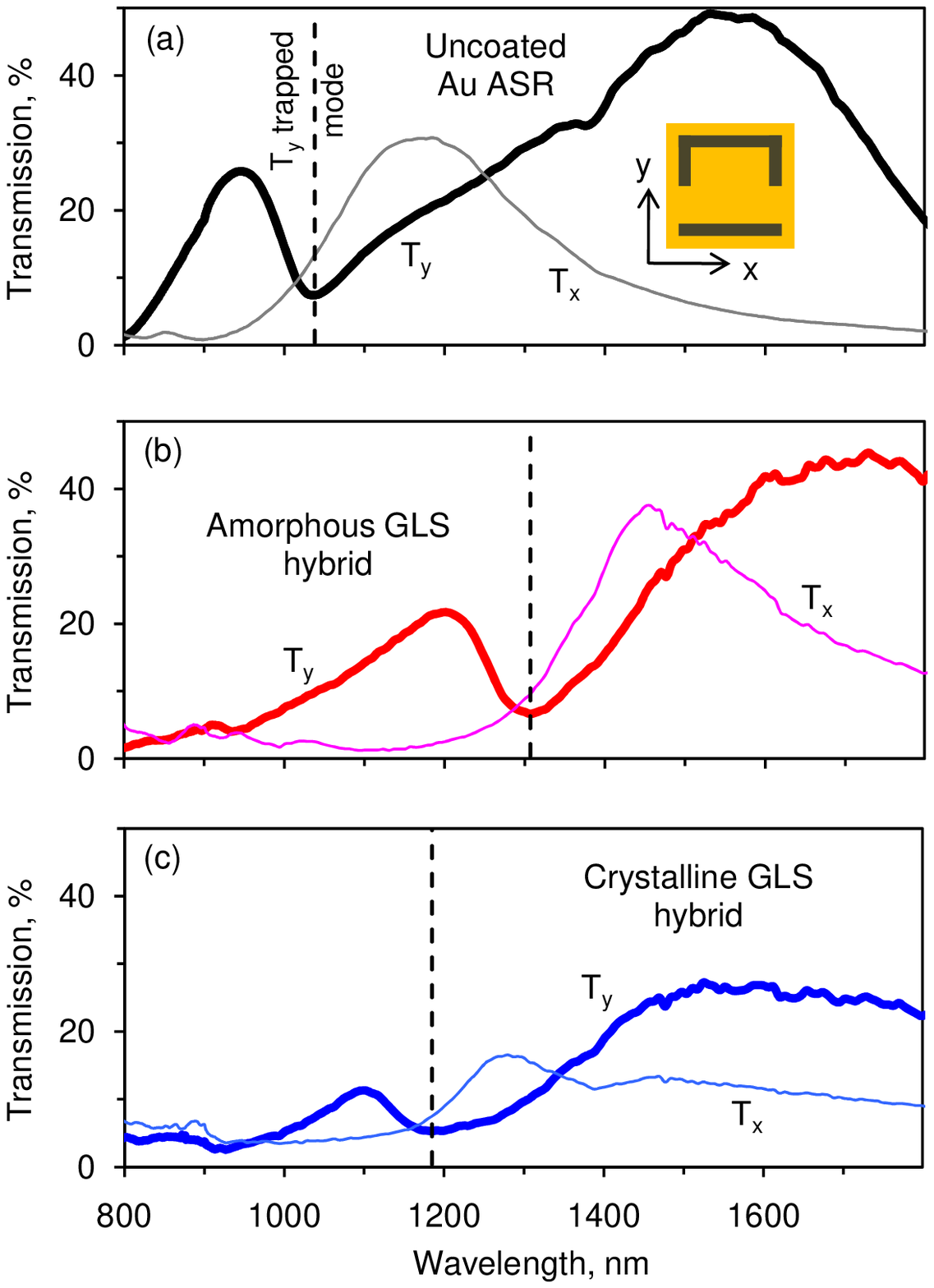}
\caption{Optical transmission spectra for: (a) the uncoated gold asymmetric split-ring metamaterial shown in Fig.~\ref{sample}a; (b and c) the same metamaterial hybridized with 200~nm gallium lanthanum sulphide chalcogenide glass film in its amorphous (b) and crystalline (c) states. Absolute transmission data are presented for incident polarizations parallel ($T_x$) and perpendicular ($T_y$) to the resonator split, as illustrated by the inset to part (a). Dashed vertical lines indicate the spectral position of the structure's trapped-mode resonance.}
\label{experiment}
\end{figure}

Optical characterizations of the metamaterial were performed using a microspectrophotometer, sampling a 24~$\mu$m$^2$ area encompassing $\sim$169 ASR unit cells. Near-infrared transmission spectra, for normally incident polarizations parallel and perpendicular to the ring resonator split direction, were recorded before GLS deposition (Fig.~\ref{experiment}a), again after deposition of amorphous-phase GLS (Fig.~\ref{experiment}b), and for a third time after electrically-induced GLS phase switching (Fig.~\ref{experiment}c).

Measurements on the uncoated gold ASR array (Fig.~\ref{experiment}a) reveal polarization-sensitive resonance features: For light polarized parallel to the resonator split direction (designated as the $x-$direction as defined in the inset to Fig.~\ref{experiment}a) there is a single broad transmission peak centered at 1170~nm. For the orthogonal y-polarization, a transmission dip associated with the excitation of the metamaterial's trapped-mode resonance is clearly seen at 1040~nm.

The presence of a 200~nm amorphous GLS film substantially changes the near-field dielectric environment of the ASRs and is found to red-shift the structure's trapped mode resonance by 270~nm to a centre wavelength of 1310~nm (Fig.~\ref{experiment}b).

Electrical phase switching of the chalcogenide film was achieved by applying 10~ms pulses of incrementally increasing voltage between the patterned gold layer and a wire probe electrode brought into contact with the top surface of the GLS~\cite{Simpson2008}. A voltage/current source measure unit (Keithley~238) was used for this purpose, which provided for real-time monitoring of the GLS film's electrical properties and thereby the facile identification of the transition point: the chalcogenide maintains a high (amorphous-state) resistivity until the applied voltage reaches approximately 45~V, at which point threshold switching occurs and through localized Joule heating it is converted to the more conductive crystalline form. This transition brings about a dramatic blue-shift of 120~nm in the spectral position of the hybrid metamaterial's trapped mode resonance (see Fig.~\ref{experiment}c) and an associated 75\% change in transmitted intensity at the crystalline-state resonance position (1190~nm). Thus, the hybrid metamaterial acts as a transmission modulator, providing a contrast ratio of 4:1 in a device structure less than one third of a wavelength thick. (The reverse transition may be induced using shorter, higher voltage pulses.)

\begin{figure}
\includegraphics*[width=\columnwidth]{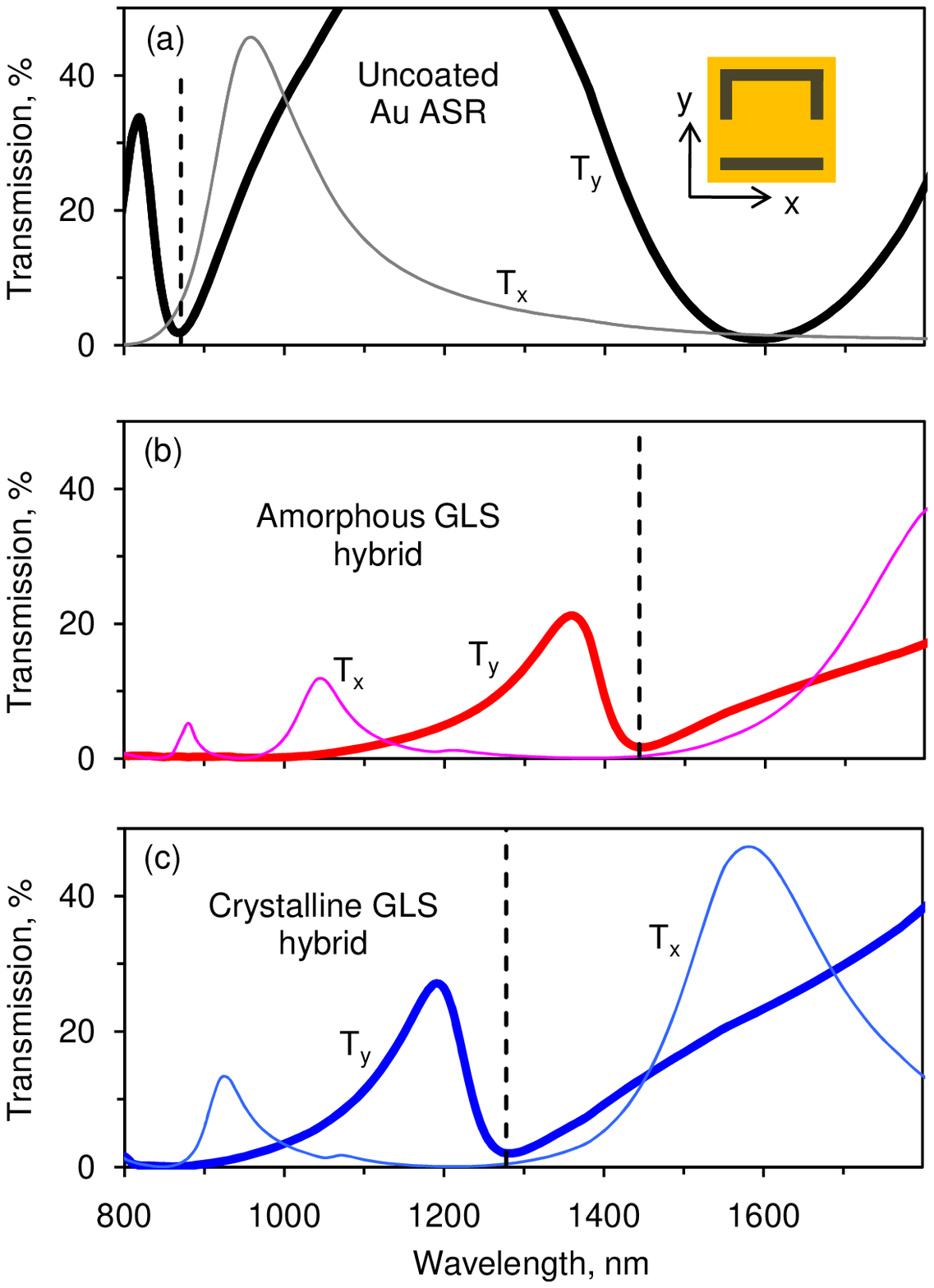}
\caption{Numerically simulated transmission spectra for (a) a 70~nm thick gold asymmetric split-ring array, with dimensional parameters as shown in the inset to Fig.~\ref{sample}a, on a 100~nm silicon nitride membrane; (b and c) the same metamaterial hybridized with 200~nm gallium lanthanum sulphide film in amorphous (b) and crystalline (c) states. Spectra are presented for $x-$ and $y-$polarizations as defined in the the inset to part (a). Dashed vertical lines indicate the spectral position of the trapped-mode resonance.}
\label{simulation}
\end{figure}

It is instructive to compare the experimental transmission spectra (Fig.~\ref{experiment}) with corresponding curves derived from a finite element numerical (Comsol Multiphysics) model of the hybrid metamaterial (Fig.~\ref{simulation}). Refractive index data for gold and silicon nitride are taken from Ref.~\cite{Palik1984} and are derived for GLS from ellipsometric measurements on separately prepared amorphous and crystalline test samples. The numerical results show a very good qualitative agreement with experiment, reproducing the form of the resonance features and the shifts associated with GLS deposition and phase switching.

Quantitative discrepancies between the two are attributed to manufacturing imperfections in ASR array fabrication (variations in line widths, depths and corner radii) and imprecise knowledge of the component refractive indices in the hybrid structure. Interestingly, the resonance positions are found to depend sensitively on the depth of GLS within the `trenches' of the negative ASR design (incomplete filling blue-shifts the resonant wavelength), whilst remaining largely insensitive to the thickness of GLS above the top surface of the gold film.

In summary, the phase-change media behind today's rewritable optical disk technologies have been shown to offer a new paradigm for \emph{active} metamaterial functionality: electrically-induced near-infrared resonance switching has been demonstrated in a photonic metamaterial hybridized with a bistable, CMOS/SOI-compatible chalcogenide glass. A 10\% shift in the resonant wavelength and a 4:1 contrast ratio in optical transmission modulation are achieved in a device structure less than one third of a wavelength thick. Comparable performance is expected in substantially thinner devices, which may be structurally engineered to operate at wavelengths throughout the visible and infrared spectral range.

\begin{acknowledgments}
This work was funded in the UK by the Engineering and Physical Sciences Research Council and in Italy under EU projects FP6-NMP-STRP 032131 `DIPNA' and FP7-NMP 229375-2 `SMD', and the Ministry of Education, University and Research project MIUR-PRIN2008.
\end{acknowledgments}


\end{document}